# The Symmetries and Redundancies of Terror: Patterns in the Dark, A study of Terrorist Network Strategy and Structure

**Philip V. Fellman,**
Southern New Hampshire University
Manchester, New Hampshire
Shirogitsune99@yahoo.com

**Mark Strathern,**
Cranfield University
Bedford, UK
m.strathern@cranfield.ac.uk

## Introduction

The importance of understanding the dreadful 9/11 attack cannot be overestimated. Krebs [Krebs 2002], has made a useful contribution using social network theory whilst Dombrowski and Carley [Dombrowski 2002] have developed a useful Bayesian technique for combining sparse and maybe unreliable informant information to build up a picture of a terrorist network. The starting point for the present analysis is the same Sydney Morning Herald article [Sydney 2001] that Valdis Krebs used in developing his now famous network analysis of the 9/11 Hijackers. However our analysis leads in a different, and complementary, direction to that uncovered by Krebs and Dombrowski. While the analysis we develop here does not provide predictive certainty, it does lead to some useful insights into the thinking, planning and organization of the terrorist networks involved in some of the worst al-Qa'ida attacks, in particular 9/11.

We approach the problem of uncovering and analyzing the structure of the 9/11 terrorist network through an exploration of the traces which any organization must inevitably leave after conducting a complex terrorist operation. We use these traces to develop a set of conclusions which are confirmed by other external data. This first round of analysis in turn forms the basis of subsequent insights and allows us to ask some key definitional questions. In addition to our own line of argument, we also provide confirming evidence for a number of our assumptions in external sources. This data supports both our methodology and the conclusions which we draw. The bulk of this evidence is simply drawn from after-the-fact public data on the 9/11 attack as well as other al-Qa'ida terrorist operations (Figure 1-below):

*Sydney Morning Herald 26th Sept 2001*

*Source: Sydney Morning Herald 26th Sept 2001 (http://old.smh.com.au/news/0109/26/world/ visited: 27/08/02)*

**The Social Network Approach**

The analysis developed here is worked up from the same publicly available data that Krebs [Krebs 2002] initially used from the Sydney Morning Herald [Sydney 2001], just fifteen days after the attack. Krebs uses this data as the start of a social network analysis.

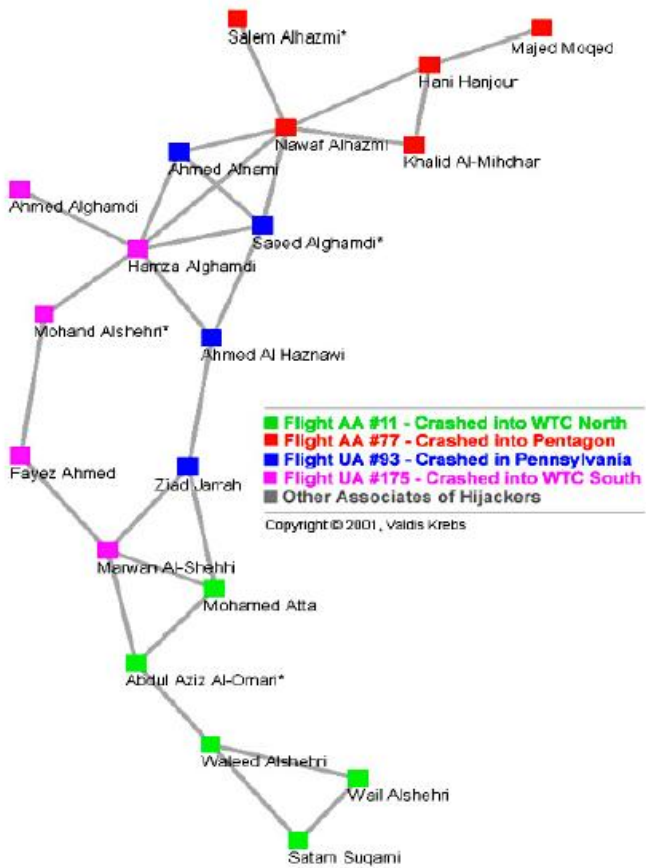

Figure 2: Valdis Krebs Social Network Diagram of the 9/11 Hijacker Network

While Krebs' work represents a landmark achievement in understanding terrorist networks, we need to keep in mind that the 9/11 terrorist network was not a normal social network. Rather, it was formally designed for a special purpose. That is, it was firstly designed to carry out a particular plan, and secondly to conceal that plan and those involved in it till after the plan had been executed. However the attacks were organized and organisation implies order and order leaves traces. We have therefore re-analysed the data using attribute mapping to find what traces of organizational order show.

**Basic Symmetries in the 9/11 Network**

Organisational order frequently appears in one of two distinct ways, as pattern (symmetry of structure) [Prigogine 1980, Simon 1957] and hierarchy (or cardinality) and both of these play a part in our enquiry. In this context, order tends to imply organisation, and frequently the reverse is also true, that organisation implies some order. If we throw a pair of dice once and they come up with a six and a one we may well think that this was the luck of the throw. If we repeat the throw a number of times and they keep on coming up with a six and a one we would presume that the dice are biased and the bias is acting as an 'organising' force pushing away from a true, random distribution towards a more ordered one consisting of mainly sixes and ones, and these become the signature of the bias.

We could also make inferences about the motivation for biasing the dice, particularly if you were at a craps table at the time! The process that we will go through in the following analysis is similar. We start with the 'assumption' that the 9/11 attacks were a single organized attack. And then we look for the traces of the order this implies. The nature of the order, and the breaks in it, will give us clues to the thinking, planning and capabilities of those involved in the operation. When we see signals with the same signature in other al-Qa'ida attacks it helps to justify the inferences that we make about the nature of the thinking and planning involved.

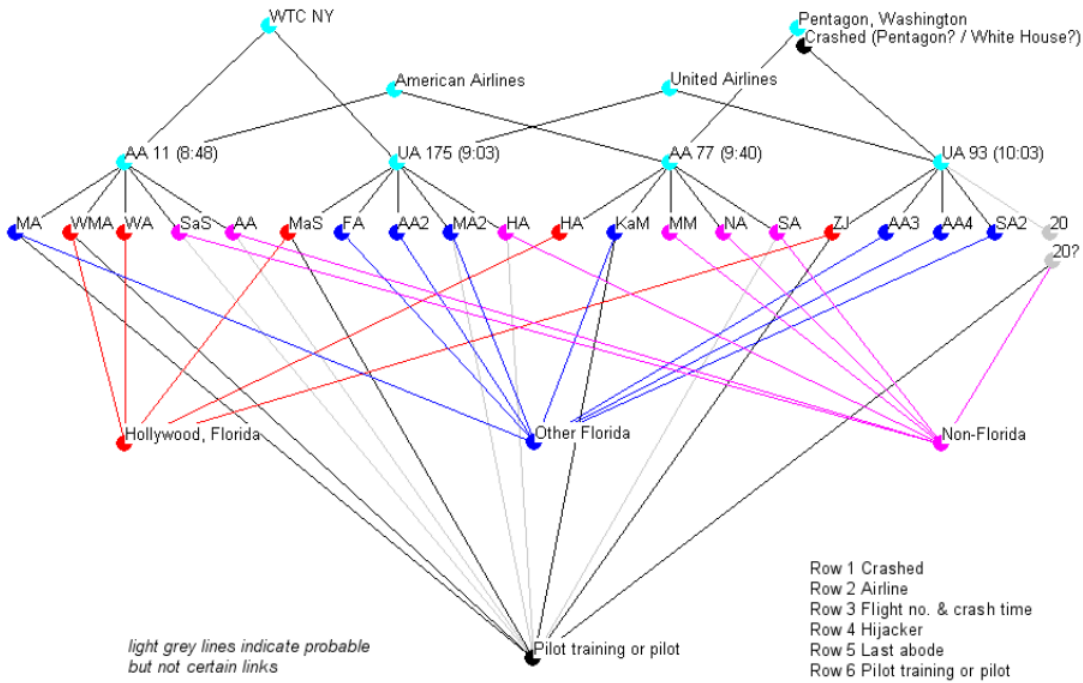

Figure 3: Symmetries and redundancies in the training of the 9/11 Hijackers

# Appendix One: Goodyear vs. Michelin – Offensive Counter Strategy
# Gary Hamel and C.K. Prahalad – Harvard Business Review, July-August 1985[1]

As a starting point, let's take a look at what drives global competition. It begins with a sequence of competitive action and reaction:

- An aggressive competitor decides to use the cash flow generated in its home market to subsidize an attack on markets of domestically oriented foreign competitors.
- The defensive competitor then retaliates—not in its home market, where the attack was staged—but in foreign markets where the aggressor company is most vulnerable.

As an example, consider the contest between Goodyear and Michelin. By today's definitions, the tire industry is not global. Most tire companies manufacture and distribute for the local market. Yet Michelin, Goodyear and Firestone became locked in a fiercely competitive—and very global battle.

In the early 1970's, Michelin used its strong European profit base to attack Goodyear's American home market. Goodyear could fight back in the United States by reducing prices, increasing advertising, or offering dealers better margins. But because Michelin would expose only a small amount of its worldwide business in the United States, it had little to lose and much to gain. Goodyear, on the other hand, would sacrifice margins in its largest market.

Goodyear ultimately struck back in Europe, throwing a wrench in Michelin's money machine. Goodyear was proposing a hostage trade. Michelin's long-term goals and resources allowed it to push ahead in the United States. But at least Goodyear slowed down the pace of Michelin's attack and forced it to recalculate the cost of market share gains in the United States. Goodyear's strategy recognized the international scope of competition and parried Michelin's thrust.

Manufacturers have played out this pattern of cross-subsidization and international retaliation in the chemical, auto, aircraft engine and computer industries. In each case, international cash flows, scale economies or homogenous markets, finally determined whether competition was global or national. The Goodyear vs. Michelin case helps to distinguish among:

- Global competition, which occurs when companies cross-subsidize national market share battles in pursuit of global brand and distribution positions.
- Global businesses, in which the minimum volume required for cost efficiency is not available in the company's home market.
- Global companies, which have distribution systems in key foreign markets that enable cross-subsidization, international retaliation, and world scale volume.

*The analogies to Al-Qaeda here should be obvious to the reader. In many ways the "battle for hearts and minds" is very similar to commercial battle for market share. One aspect of this which policy makers would be wise to heed comes from Michael Porter's "Cost-Focus" matrix, which is used to differentiate between high value added specialty or niche-market products, low value added mass produced products (both of which represent successful strategies) and low value added/high cost companies and industries (Porter uses the garment business in New York as his example) with the occasional low cost/high value added corporation like Microsoft which earns extraordinary returns (which could be argued is a consequence of Microsoft's monopoly position with respect to many computing goods and services)[2]. However, as the following appendix, excerpted from a longer RAND corporation report shows, the analogies are not just conceptual metaphors, but correspond to a "learning organization" with flexible structure.*

---

[1] Excerpted from Hamel, Gary and Prahalad, C.K. (1985) "Do you really have a global strategy', Harvard Business Review, July-August, 1985.
[2] Porter, Michael (1990) "The Competitive of Nations", Harvard Business Review, March-April, 1990.

# Appendix II: Redefining Counterterrorism-The Terrorist Leader as CEO
# By Bruce Hoffman, Rand Corporation[3]

Killing Usama bin Laden will not quash the terrorist threat from al Qaeda, because the group sees the war it started as an epic struggle lasting years if not decades. The group has shown itself to have a deeper "bench" than was previously thought and to have some form of "corporate succession" plan. In fact, the closest organizational relative to al Qaeda is perhaps a private multinational corporation. And bin Laden himself is perhaps best viewed as a terrorist CEO.

He has applied business administration and modern management techniques learned both at the university and in the family's construction business to the running of a transnational terrorist organization. He obtained a degree in economics and public administration in 1981 from Saudi Arabia's prestigious King Abdul Aziz University. He then cut his teeth in the family business, honing the management and organizational skills that later enabled him to transform al Qaeda into the world's preeminent terrorist movement.

He has implemented for al Qaeda the same type of effective organizational framework adopted by many corporate executives throughout much of the industrialized world over the past decade. Just as large, multinational business conglomerates moved during the 1990s to flatter, networked structures, bin Laden did the same with al Qaeda.

He defined a flexible strategy for the group that functions at multiple levels, using both top-down and bottom-up approaches. On the one hand, he has functioned like the president or CEO of a large multinational corporation by defining specific goals, issuing orders, and ensuring their implementation. This function applies mostly to the al Qaeda" spectaculars"—those high-visibility, usually high-value, and high-casualty operations like 9/11, the attack on the USS Cole, and the 1998 east Africa embassy bombings.

On the other hand, he has operated as a venture capitalist by soliciting ideas from below, by encouraging creative approaches and out-of-the-box thinking, and by providing funding to those proposals he finds promising. Several attacks by groups affiliated with al Qaeda attest to this approach. The attacks include those staged by Jemaah Islamiyah in Bali in October 2002 and Jakarta in August 2003; by al-Assiriyat al-Moustaqim in Morocco in May 2003; and by the Islamic Great Eastern Raiders Front in Turkey in November 2003.

Al Qaeda deliberately has no single, set modus operandi—which makes the group all the more resilient and formidable. Instead, bin Laden built a movement that actively encourages subsidiary groups fighting under the corporate banner to mix and match approaches, employing different tactics and varying means of attack and operational styles in a number of locales.

Even in the post 9/11 era, when al Qaeda has been relentlessly tracked, harassed, and weakened, the corporate succession plan seems to have functioned. The group appears to retain at least some depth in numbers as evidenced by its replenishment abilities to produce successor echelons for the mid-level operational commanders who have been killed or captured.

---

[3] http://www.rand.org/publications/randreview/issues/spring2004/ceo.html

# Appendix III: Network Structure and Organizational Typology

*In a joint paper presented at the NATO conference on Central Asia last year, we and our colleagues argued that in terrorist networks which have a widely distributed structure, like the Al Qaeda allies in Indonesia or the Philippines, gains might be made against such organizations by forcing them to "over-compartment" there operations and their cells, thus limiting the ability of the leadership to propagate messages across the "network".[4] We drew largely on the work of Moody and White in constructing a social network model of Al-Qaeda as is reproduced below:[5]*

Moody and White provide an expansion of the social solidarity concept and the understanding of linkages between members of a community, the changing interconnections and the impact on node connectivity in "Social Cohesion and Embeddedness".[6] They argue that the defining characteristic of a strongly cohesive group is that "it has a status beyond any individual group member". The authors define structural cohesion as "the minimum number of actors who, if removed from a group, would disconnect the group", leading to hierarchically nested groups, where highly cohesive groups are embedded within less cohesive groups. Thus, cohesions is an emergent property of the relational pattern that holds a group together.

As the dynamical process of group development unfolds, typically a weak form of structural cohesion begins to emerge as collections of unrelated individuals begin connecting through a single path which reflects new relationships. As additional relations form among previously connected pairs of individuals, multiple paths through the group develop, increasing the community's ability to "hold together".

In situations where relations revolve around a leader, the group is often described as "notoriously fragile", illustrating the fact that *increasing relational volume thru a single individual does not necessarily promote cohesiveness* . Nevertheless, groups with an all-in-one relational organization such as *terrorist networks, may be stable and robust to disruptions if "extraordinary efforts" are put into maintaining their weak relational structure*. The spoke-and-hub configuration of these networks thrives on the lack of knowledge that each particular node has about the organization as a whole, a captured or destroyed link in the network does not put the organization at risk. The stability of such groups depends on the ability to keep the hub hidden, because *the hub then becomes the entire group's fundamental structural weakness*.

Weakly cohesive organizations also promote segmentation into structures that are only minimally connected to the rest of the group, leading to schisms and factions. These organizations are also easily disrupted by individuals leaving the group. *Usually, individuals whose removal would disconnect the group are those in control of the flow of resources in the network*.

On the contrary, *collectivities that do not depend on individual actors are less easily segmented*. These highly cohesive groups benefit form the existence of multiple paths and sets of alternative linkages, with no individual or minority within the group exercising control over resources. The "multiple connectivity" is thus the essential feature of the strong structurally cohesive organizations.

An interesting characteristic of such highly cohesive networks (**HCN's**) is that they are characterized by a reduction in the power provided by structural holes, such that the ability of any individual to have power within the setting is limited as connectivity increases.[7] For a structurally cohesive group, the information transmission increases with each additional independent path in the network, which may infer that high connectivity leads to more reliability as information is combined from independent multiple sources. "local pockets of high connectivity" can act as "amplifying substations" of information and/or resources. Moody and White relate this operationalization to the actors' relative involvement depth in social relations, as defined by the concept of embeddedness.:

> "If cohesive groups are nested within each other, then each successive group is more deeply embedded within the network. As such, one aspect of embeddedness- the depth of involvement in a relational structure- is captures by the extent to which a group is nested within the relational structure."

---

[4] Fellman, Philip V.; Wright, Roxana and Sawyer, David (2003) "Modeling Terrorist Networks - Complex Systems and First Principles of Counter-Intelligence", NATO and Central Asia: Enlargement, Civil – Military Relations, and Security Kazach American University/North Atlantic Treaty Organization (NATO) May 14-16, 2003.

[5] James Moody, Douglas R. White, "Social Cohesion and Embeddedness: A Hierarchical Conception of Social Groups, Santa Fe Institute Working Papers, 00-08-049, http://www.santafe.edu/sfi/publications/Working-Papers/00-08-049.pdf

[6] James Moody, Douglas R. White, "Social Cohesion and Embeddedness: A Hierarchical Conception of Social Groups, Santa Fe Institute Working Papers, 00-08-049, http://www.santafe.edu/sfi/publications/Working-Papers/00-08-049.pdf

[7] In terms of strategy, destabilizing this kind of network means pressuring the group to increase its recruitment and raise its connectivity as opposed to the previously discussed strategy of forced over-compartmentation. Induced excess connectivity represents a different kind of complexity overload.

In a companion paper to Moody and White's "Social Cohesion and Embeddedness", White and Harary, in distinguish between the adhesion concept related to the attractive or charismatic qualities of leaders (or attractions to their followers) that create weaker or stronger many-to-one ties or commitments, and the cohesion defined by the many-to-many ties among individuals, as they form into clusters.[8] The authors reiterate the intuitive aspects of the cohesion's definition: a group is cohesive to the extent that the social relations of its members are resistant to the group being pulled apart, and a group is cohesive to the extent that the multiple social relations of its members pull it together. In revisiting the idea that ***minimal cohesion occurs in social networks with a strong group leader or popular figure***, White and Harary introduce the concept of ***"adherents"*** of a social group to specify ***"the many-to-one commitments of individuals to the group itself or to its leadership"***.

> "***What holds the group together where this is the major factor in group solidarity is the strength of adhesion of members to the leader, not the cohesiveness of group members in terms of social ties amongst themselves***. The model of "adhesion" rather than cohesion might apply to the case of a purely vertical bureaucracy where there are no lateral ties. "

As a general definition, a group is adhesive to the extent that the social relations of its members are pairwise-resistant to being pulled apart.[9] Another element of group robustness is the redundancy of connections:[10]

> The level of cohesion is higher when members of a group are connected as opposed to disconnected, and further, when the group and its actors are not only connected but also have redundancies in their interconnections. The higher the redundancies of independent connections between pairs of nodes, the higher the cohesion, and the more social circles in which any pair of persons is contained

The important consideration for counter-intelligence here is that the higher the level of redundancy, the more likely the existence of the group is to be revealed and the easier it is to create a map of social network relationships. A major part of the successful exploitation of this group characteristic is another of the three basic counter-intelligence principles—coverage. Good coverage will yield good observations from which good social network maps can be derived. The caution here, as we have already noted, is like many other HUMINT activities, good coverage is not possible to achieve solely by satellite reconnaissance or any other national technical means (NTM). Of course, once target individuals have been identified, dedicated remote sensing technology can, in fact, be a very helpful adjunct to the processes of coverage, compartmentation and penetration.

Again, just as technology cannot substitute for on the ground coverage, effectively covering the broad range of terrorist groups in the world today is also not possible with case officer staffing at the 15% level in embassies. It is far less achievable without a highly trained group of multi-lingual, multi-talented individuals who are both willing and able to work successfully for extended periods of time in local environments as local citizens with no official cover or identifying features as "foreign" nationals (NOC's). One thing that mid-range analysis tells any intelligent decision maker is that you can have all the science in the world at your finger tips, but without an effective human organization capable of learning and carrying out the missions which that science prescribes you are paralyzed.  Remember, the 9/11 hijackers took those planes with box-cutters and pocket knives, not AK-47's, machine guns, intelligent weapons or recoilless rifles and TOW launchers.

---

[8] The Cohesiveness of Blocks in Social Networks: Node Connectivity and Conditional Density. Submitted to Sociological Methodology 2001. http://www.santafe.edu/files/workshops/dynamics/sm-wh8a.pdf
[9] The concept of cohesion is formalized through the use of graph theory. The graph is defined that the vertices represent the set of individuals in the network, and the edges are the relations among actors defined as paired sets. The subsets of nodes that link non-adjacent vertices will disconnect actors if removed. Any such set of nodes is called an (i, j) cut-set if every path connecting i and j passes through at least one node of the set . The "cut-set resistance to being pulled apart" criterion and the multiple independent paths "held together" criterion of cohesion are formally equivalent in this formal specification. This kind of graph, if constructed with complete information, also provides a predictive mechanism for exactly which nodes need to be removed in order to remove the possibility of signals propagating through the system.
[10] Ibid. No. 5

*This excerpt highlights the importance of redundancy in distributed organizations.  Given the kind of group structure which was used to execute the 9/11 attacks, redundancy suddenly takes on a far more significant meaning, indeed it may even be one of the central, defining characteristics of the group.*

# Appendix IV: Adaptive Dynamics, Coevolution and Complexity Catastrophe

*Another important, recent approach to modeling organizations has come from complexity theory, perhaps most clearly explained by* **Emergence, Complexity and Organization***'s founding editor, Michael Lissack.[11] Where this approach can be used to provide more concrete results is in the dynamic fitness landscape model developed originally by Sewall Wright in the 1930's and subsequently revived and greatly expanded by Stuart Kauffman.[12] Bill McKelvey has done a very precise study using statistical mechanics to decompose Michael Porter's value chain of a firms activities into scalar values which can then be plotted against the values of rivals. In a paper given at the 2004 International Conference on Complex Systems, several of us presented papers which examined a variety of organizations using Kauffman's model.[13] In describing Bill McKelvey's approach, which bears very strongly on counter-terrorism methods like forced overcompartmentation, we noted:*

While the concept of the fitness landscape is not new[14] Kauffman was the first to place fitness landscapes in a dynamic setting using Boolean networks.[15] Kauffman explains rugged fitness landscapes as a natural outcome of adaptive evolution:[16]

> Adaptive evolution occurs largely by the successive accumulation of minor variations in phenotype. The simple example of computer programs makes it clear that not all complex systems are graced with the property that a minor change in systemic structure typically leads to minor changes in system behavior. In short, as their internal structure is modified, some systems change behavior relatively smoothly and some relatively radically. Thus we confront the question of whether selective evolution is able to "tune" the structure of complex systems so that they evolve readily.

The mechanism by which biological systems tune is the result of standard statistical mechanics, whereby ensembles move randomly through all phase spaces of the system over time. With respect to the random array of ensemble positions, we might think of economic or business systems which occupy nearby positions in a large market with many securities (a "densely packed" state space). Borrowing from random walk theory we could view the ability of a security to earn economic returns as the measure of its fitness. While this is a very simple metaphor it helps to tie the conceptual foundations of economics and finance to evolutionary biology in the sense that the elements of both systems can rest in peaks or valleys (in finance this would be equivalent to Jensen's $\alpha$, which represents the degree to which management (and by logical inference, corporate structure) either creates or destroys value in a firm. The peaks can be either local maxima or global maxima. In highly efficient markets stocks would reach their equilibrium prices rather rapidly and the fitness landscape would be highly correlated.[17]

In less efficient markets, for example markets like the long distance commodities trades we frequently see in international business, where there are significant information asymmetries and market inefficiencies, the landscape would tend to be "rugged" and multi-peaked. In such a situation, there are opportunities for large arbitrage (similar to the biologist's "punctuated equilibrium").[18] In the context of a dynamic landscape[19] where multiple entities interact and both their presence and their interactions affect the structure of the landscape, Kauffman argues:[20]

---

[11] Lissack, Michael, (1996) "Chaos and Complexity: What Does That Have to Do with Knowledge Management?", in Knowledge Management: Organization, Competence and Methodology, ed. J. F. Schreinemakers, Ergon Verlog 1: 62-81 (Wurzburg: 1996)

[12] Kauffman, Stuart (1993), The Origins of Order, Oxford University Press (Oxford and New York: 1993)

[13] Fellman, Philip V.; Wright, Roxana; Post, Jonathan Vos; and Dasari, Usha (2004), "Adaptation and Coevolution on an Emergent Global Competitive Landscape", paper presented at the International Conference on Complex Systems, Boston, May 16-21, 2004.

[14] Wright, S. (A) "Evolution in Mendelian populations", *Genetics*, 16 97–159, 1931. (B) "The roles of mutation, inbreeding, crossbreeding and selection in evolution", .Proceedings Sixth International Congress on Genetics 1 356–366 (1932)

[15] Kauffman, S. A. (1969) Metabolic Stability and Epigenesis in Randomly Constructed Genetic Nets. Journal of Theoretical Biology, 22, pp. 437-467.

[16] Ibid. No. 6

[17] For a more complete description see Smith, Eric; Farmer, J. Doyne; Gillemot, Laszlo; and Krishnamurthy, Supriya "Statistical Theory of the Continuous Double Auction", Santa Fe Institute Working Papers, October 20, 2002 http://www.santafe.edu/sfi/publications/Working-Papers/02-10-057.pdf

[18] See Aoki, Masahiko, Subjective Game Models and the Mechanism of Institutional Change, Santa Fe Institute Workshop, http://www.santafe.edu/files/workshops/bowles/aokiIX.PDF

[19] Those unfamiliar with the mechanics of Boolean networks can obtain a very simple, straightforward explanation with examples from Torsten Reil's "An Introduction to Complex Systems", Department of Zoology, Oxford University at

> The character of adaptive evolution depends on the structure of such fitness landscapes. Most critically, we shall find that, as the complexity of the system under selection increases, selection is progressively less able to alter the properties of the system. Thus, even in the presence of continuing selection, complex systems tend to remain typical members of the ensemble of possibilities from which they are drawn…Thus, if selection, when operating on complex systems which spontaneously exhibit profound order, is unable to avoid that spontaneous order, that order will *shine through*. In short, this theme, central to our concerns, states that much of the order in organisms may be spontaneous. Rather than reflecting selection's successes, such order may, remarkably reflect selection's failure.

Two important points follow from Kauffman's model. The first is that economic theories particularly equilibrium-oriented theories which model performance as the outcome of selection may have built their structure on an internally contradictory foundation from the ground up.[21] The second point is more technical and has to do with some of the formal mathematical properties NK Boolean networks as well as the mathematical limitations that coevolving populations impose on fitness landscapes. Kauffman's model treats adaptive evolution as a search process.[22]

> Adaptive evolution is a search process—driven by mutation, recombination and selection—on fixed or deforming fitness landscapes, An adapting population flows over the landscape under these forces. The structure of such landscapes, smooth or rugged, governs both the evolvability of populations and the sustained fitness of their members. The structure of fitness landscapes inevitably imposes limitations on adaptive search. On smooth landscapes and a fixed population size and mutation rate, as the complexity of the entities under selection increases an error threshold is reached…the population "melts" from the adaptive peaks and flows across vast reaches of genotype space among near neutral mutants. Conversely on sufficiently rugged landscapes, the evolutionary process becomes trapped in very small regions of genotype space…the results of this chapter suffice to say that selection can be unable to avoid spontaneous order. The limitations on selection arise because of two inexorable complexity catastrophes…each arises where the other does not. One on rugged landscapes, trapping adaptive walks, the other on smooth landscapes where selection becomes too weak to hold adapting populations in small untypical regions of the space of possibilities. Between them, selection is sorely pressed to escape the typical features of the system on which it operates.

Kauffman covers a great deal of ground here, not all of which can be fully recapitulated in a brief paper. Again, Kauffman's Type II complexity catastrophe (which takes place on smooth or highly correlated landscapes) bears a strong resemblance to what Farmer describes as "first order efficiency" in financial markets with very thin arbitrage margins for technical traders.[23] To earn "extraordinary returns" traders are forced to undertake "wide searches", equivalent to highly leveraged positions.

Kauffman's Type I complexity catastrophe is the result of the fundamental structure of NK Boolean fitness landscapes. In a dynamic fitness landscape, as complexity increases, the heights of accessible peaks fall towards mean fitness. Again, in terms of financial markets we might compare this to the fact that over long periods of time the betas of all stocks in the market have a tendency to drift towards 1.[24]

Bill McKelvey takes the fitness landscape model well past being a metaphor for market evolution and decomposes Michael Porter's value chain into value chain competencies as the parts of firms (the N of the NK network) and K as the connections of between the parts within a firm and the parts of the opponents. He models firm behavior using a heterogeneous[25] stochastic microagent assumption, which then allows him

---

http://users.ox.ac.uk/~quee0818/complexity/complexity.html Vladimir Redko provides a slightly more mathematical treatment at http://pespmc1.vub.ac.be/BOOLNETW.html

[20] Ibid. No. 6

[21] See, for example, Reaume, David "Walras, complexity, and Post Walrasian Macroeconomics," Collander David, Ed. <u>Beyond Microfoundations: Post Walrasian Macroeconomics</u>, Cambridge University Press 1996

[22] The search property of organisms seeking maximal fitness levels allows one to draw some additional, strong conclusions about both organisms and landscapes.

[23] Ibid. No. 12, See also Zovko, Ilija and Farmer, J. Doyne, "The Power of Patience: A behavioral regularity in limit order placement", Santa Fe Institute Working Papers, 02-06-27; Daniels, Marcus J.; Iori, Giulia; and Farmer, J. Doyne, "Demand Storage, Liquidity, and Price Volatility" Santa Fe Institute Working Papers, 02-01-001; and Farmer, J. Doyne and Joshi, Shareen, "The Price Dynamics of Common Trading Strategies", in the Journal of Economic Behavior and Organization October 30, 2001.

[24] See Theobald, Michael and Yallup, Peter "Determining the Security Speed of Adjustment Coefficients", http://bss2.bham.ac.uk/business/papers/Theocoeff.pdf , May 2001. See also Blake, David; Lehmann, Bruce N. and Timmerman, Allan "Asset Allocation Dynamics and Pension Fund Performance", November, 1998
http://econ.ucsd.edu/~atimmerm/pensrev2.pdf

[25] See (A) Farmer, J. Doyne "Physicists Attempt to Climb the Ivory Towers of Finance", Computing in Science and Engineering, November-December, 1999 for an explanation of agent-based models. See also W. Brian Arthur's classic game theoretic treatment of an interaction of agents which has no equilibrium and no possibility of satisfying every player either prospectively or retrospectively,

to test four basic concepts: "(1) What intrafirm levels of integrative complexity affect competitive advantage? (2) What levels of integrative complexity influence how rapidly firms in coevolutionary groups reach equilibrium fitness levels, if they do so? (3) What complexity factors might affect the competitive advantage (or height) of fitness levels? and (4) What levels of integrative complexity might affect the overall adaptive success of firms comprising a coevolving system?"[26]

The formal structure of McKelvey's question is every bit as important as the answers, some of which are provided in his work and some which are explained by the Windrum-Birchenhall model of technological diffusion which we discuss next. McKelvey's work serves two important purposes. First, he develops a metric from Porter's value chain which allows him to calculate actual scalar or vector values for different NK structures. This also allows him to calculate fitness levels in terms of Nash Equilibria which vary directly as a function of N and K.[27] Secondly, the values he derives from Porter's value chain allow him to run simulations which specify under what conditions firms should either increase or decrease their internal and external connectivity. While in many ways McKelvey is really just taking the first cut at the problem, he is doing so with rigorous statistical mechanics, so that even where his conclusions are not immediately overwhelming, they are profoundly accurate. Even though we might not expect most firms to start using Kauffman's model in the immediate future (Kauffman did run a joint venture with Ernst and Young for several years, during which his primary purpose was to organize conferences and explain to clients the value and insights of his model)[28] McKelvey's advice, ***In general the simulations indicate that keeping one's internal and external coevolutionary interdependencies just below that of opponents*** can be expected to hold true for quite some time.

From our perspective, as we see terrorist organizations evolve over time to become increasingly sophisticated, we might consider two relevant results from McKelvey's research. First, counter-terrorist organizations can become so highly compartmented that the suffer diseconomies of scope. One has merely to think of the ultra secret, NSA Decrypt vault run at TRW, as portrayed in the film "The Falcon and the

---

the El Farol Bar Game, in (B) "Inductive Reasoning and Bounded Rationality", American Economic Review, May 1994, Vol. 84, No. 2, pp. 406-411. Arthur also deals with the stochastic microagent assumption in(C) Arthur, W. Brian, "Learning and Adaptive Economic Behavior: Designing Economic Agents that Act Like Human Agents: A Behavioral Approach to Bounded Rationality", American Economic Review, Vol. 81, No. 2, May 1991, pp. 353-359.

[26] McKelvey, Bill "Avoiding Complexity Catastrophe in Coevolutionary Pockets: Strategies for Rugged Landscapes", Organization Science, Vol. 10, No. 3, May–June 1999. McKelvey explains his use of Kauffman's model by noting that "The biologist Stuart Kauffman suggests a theory of complexity catastrophe offering universal principles explaining phenomena normally attributed to Darwinian natural selection theory. Kauffman's complexity theory seems to apply equally well to firms in coevolutionary

pockets. Based on complexity theory, four kinds of complexity are identified. Kauffman's "NK[C] model" is positioned "at the edge of chaos" between complexity driven by "Newtonian" simple rules and rule driven deterministic chaos. Kauffman's insight, which is the basis of the findings in this paper, is that complexity is both a consequence and a cause. Multicoevolutionary complexity in firms is defined by moving natural selection processes inside firms and down to a "parts" level of analysis, in this instance Porter's value chain level, to focus on microstate activities by agents.

[27] Ibid., "Kauffman argues that his "NK[C] Boolean game" model behaves like Boolean networks when agent outcomes are limited to 0 or 1, the K number of interdependencies is taken as the number of inputs, and Nash equilibria in N person games are taken as equivalent to agents being trapped on local optima. In the NK Boolean game, fitness yields are assigned to the 0 or 1 actions by drawing from a uniform distribution ranging from 0.0 to 1.0. The K epistatic interdependencies that inhibit fitness yields from an agent's actions are drawn from a fitness table in which fitness levels of each "one-change" nearest neighbor are assigned by drawing from a uniform distribution also ranging from 0.0 to 1.0. Kauffman points out that the complexity tuning effect occurs when increasing K reduces the height of local optima while also increasing their number. Thus, high K leads to complexity catastrophe. In describing how K and C effects enter into the model, Kauffman says:

. . . [F]or each of the N traits in species 2, the model will assign a random fitness between 0.0 and 1.0 for each combination of the K traits internal to species 2, together with all combinations of C traits in species 1. In short, we expand the random fitness table for each trait in species 2 such that the trait looks at its K internal epistatic inputs and also at the C external epistatic inputs from species 1 (Kauffman 1993, p. 244).

One might conclude from this that K and C are combined into one overall moderating effect on the fitness yield from an agent's choice to adopt a higher fitness from a nearest-neighbor. Results of the models indicate otherwise.

As Kauffman points out (pp. 249, 254), the speed at which agents encounter Nash equilibria increases with K, and decreases as C and S increase. Thus, in these models K acts as a complexity "forcing" effect in speeding up the process of reaching stable Nash equilibria, whereas C acts as an "antiforcing" effect, as does S. Presumably K effects are averaged as per the basic NK model, leaving C and S effects (S multiplies the C effects) to modify fitness yields on an agent's actions independently of K effects. The consequence is that increasing K tunes the landscape toward more ruggedness (increased numbers of less fit local optima), and increases the likelihood of agents being marooned on local optima. But increasing C and/or S prevents achieving Nash equilibrium by prolonging the "coupled dancing," as Kauffman (1993, p. 243) calls it, in which opponents keep altering each other's landscapes, keep the fitness search going, and thereby prevent stability—the more opponents there are, the more the instability persists.

[28] See Meyer, Chris "What's Under the Hood: A Layman's Guide to the Real Science", Ernst and Young/Cap Gemini Center for Business Innovation, http://www.cbi.cgey.com/events/pubconf/1996-07-19/proceedings/chapter%2010.pdf

Snowman"[29] (based upon a true story chronicled by Robert Lindsey[30]) where the document shredder was primarily used to mix "Margaritas". Bureaucracy has many dysfunctional forms, and historical evidence suggests that a certain degree of "lean-ness" may serve better than a massive bureaucratic structure.[31]

By the same token, a basic policy recommendation should be that once we can get an accurate picture of a terrorist network[32] it might well be more useful to force an organization to "load up" (i.e., demand more funding, consume more resources, etc.) in order to bring about the kind of breakdown which McKelvey characterizes as a complexity catastrophe. Additional work by Carley et al, has examined the impact or cognitive load, and as a corollary, cognitive overload as a mechanism for impeding the functioning of terrorist networks.[33]

---

[29] See http://www.imdb.com/title/tt0087231/ for details.
[30] Lindsey, Robert (1979) The Falcon and the Snowman: A True Story of Friendship and Espionage, Simon and Schuster (New York: 1979).
[31] See Knott, Jack and Miller, Gary (1988) Reforming Bureaucracy, Prentice-Hall (Englewood Cliffs: 1988).
[32] Several tools for dynamic network analysis and for mapping covert networks with incomplete information are in the process of development under Kathleen Carley's supervision at Carnegie Mellon University. See, for example, Carley, Kathleen; Dombrowski, Matthew; Tsvetovat, Max; Reminga, Jeffrey; and Kaneva, Natasha (2003) "Destabilizing Dynamic Covert Networks", Institute for Software Research International, http://www.casos.cs.cmu.edu/publications/resources_others/a2c2_carley_2003_destabilizing.pdf
[33] See Carley, Kathleen; Lee, Ju-Sung and Krackhardt, David (2001), "Destabilizing Networks", CONNECTIONS 24(3): 31-34 © 2001 INSNA.